\documentclass[letterpaper, 10 pt, conference]{ieeeconf}  

\IEEEoverridecommandlockouts                              

\usepackage{graphics} 
\usepackage{epsfig} 
\usepackage{cite}
\usepackage[export]{adjustbox}

\title{\LARGE \bf
Towards RehabCoach: Design and Preliminary Evaluation of a Conversational Agent Supporting Unsupervised Therapy after Stroke
}

\author{Giada Devittori$^{1}$, Mehdi Akeddar$^{1}$, Alexandra Retevoi$^{1}$, Fabian Schneider$^{2}$, Viktoria Cvetkova$^{3}$, Daria\\ Dinacci$^{3}$, Antonella Califfi$^{3}$, Paolo Rossi$^{3}$, Claudio Petrillo$^{3}$, Tobias Kowatsch$^{2,4,5,6}$ and Olivier Lambercy$^{1,6}$ 
\thanks{This work was supported by the Competence Centre for Rehabilitation Engineering and Science (RESC) of ETH Zurich, and the National Research Foundation, Prime Minister’s Office, Singapore under its Campus for Research Excellence and Technological Enterprise (CREATE) program.}
\thanks{$^{1}$Rehabilitation Engineering Laboratory, ETH Zurich, Switzerland
        {\tt\small giada.devittori@hest.ethz.ch}}%
\thanks{$^{2}$Department of Technology, Management, and Economics, ETH Zurich, Switzerland}%
\thanks{$^{3}$Clinica Hildebrand Centro di riabilitazione Brissago, Switzerland}%
\thanks{$^{4}$Institute for Implementation Science in Health Care, University of Zurich, Switzerland}%
\thanks{$^{5}$School of Medicine, University of St. Gallen, Switzerland}%
\thanks{$^{6}$Future Health Technologies program, Singapore-ETH Centre, Campus for Research Excellence and Technological Enterprise (CREATE), Singapore}%
}

\begin{document}

\maketitle
\thispagestyle{empty}
\pagestyle{empty}

\begin{abstract}

Unsupervised therapy after stroke is a promising way to boost therapy dose without significantly increasing the workload on healthcare professionals. However, it raises important challenges, such as lower adherence to therapy in the absence of social interaction with therapists. We present the initial prototype of RehabCoach, a novel smartphone-based app with conversational agent to support unsupervised therapy. RehabCoach is designed to increase patients’ engagement and adherence to therapy and to provide information (e.g., about stroke, health) in an interactive and user-friendly manner. We report on the design and usability evaluation of the first prototype of RehabCoach, assessed by four stroke patients and five healthcare professionals, who interacted with the app in a single testing session. Task completion time and success rates were measured for 15 representative tasks, and participants assessed usability via questionnaires and a semi-structured interview. Results show that it was feasible for stroke patients to successfully interact with RehabCoach (task success $\geq$93 \%) without requiring extensive training. Participants positively rated the usability of RehabCoach (mean mHealth App Usability Questionnaire score: 1.3 for primary users, 1.4 for healthcare professionals, on a scale from 1 (positive evaluation) to 7). The feedback collected in this work opens the door to further enhance RehabCoach as an interactive digital tool to support unsupervised rehabilitation.

\end{abstract}

\section{INTRODUCTION}

People after a stroke often do not fully recover upper limb function \cite{Lawrence2001}, which negatively impacts independence and quality of life. Increasing the dose (i.e., task repetitions and therapy time) of upper limb therapy may enhance functional outcomes or prevent the long-term loss of the improvements gained during the early rehabilitation phase \cite{McCabe2015ComparisonTrial,Ward2019IntensiveProgramme}. However, the current rehabilitation model, mainly based on one-to-one therapy sessions, and the limited resources available (e.g., low therapist-to-patient ratio \cite{Jesus2017HumanCentury}) make increasing therapy dose challenging.

Unsupervised therapy, defined as patients training without any direct external supervision, might be a way to increase dose without further straining the healthcare system. Especially when applied in the home setting, unsupervised therapy could help decrease the reliance on hospital stays and visits, thus ensuring access to rehabilitation services even when the healthcare system is under pressure, as exemplified during the COVID-19 pandemic \cite{Bersano2020StrokeCountries, AguiardeSousa2020MaintainingOrganisation}. Furthermore, studies have shown that unsupervised therapy can be comparable to standard care \cite{Olney2006}.

The literature describes a wide range of tools that may support unsupervised therapy, such as booklets of exercises \cite{Zondervan2016Home-basedProgram}, virtual reality systems \cite{Thielbar2020Home-basedTrial, Wittmann2016Self-directedSystem}, and actuated robots \cite{Ranzani2023DesignFunction, Devittori2023}. However, unsupervised therapy raises important challenges, such as low engagement and motivation to train, poor adherence to therapy schedule, as well as lack of social interaction and feedback on progress \cite{Chen2019Home-basedReview, Gelineau2022ComplianceReview, Olney2006}. Attempts to address these challenges include weekly communication with a therapist \cite{Wolf2015TheRehabilitation} or the implementation of multiuser therapy exercises \cite{Thielbar2020Home-basedTrial}, which can increase adherence but rely on the presence of external persons. Integrating a gaming environment into rehabilitation may also increase motivation \cite{Karamians2020EffectivenessMeta-analysis}, but it does not solve the issue of lack of social interaction. 

Digital interventions may play a key role in addressing these gaps \cite{Lang2022}. To explore this, we are proposing RehabCoach, a novel chatbot-based mobile application designed to complement and support existing methods for unsupervised therapy. RehabCoach is based on the MobileCoach framework \cite{MobileCoach, Filler2015MobileCoach:Context, Kowatsch2017DesignMobileCoach}, an open-source platform for health interventions that relies on the establishment of a personal relationship between the user and a digital coach, mainly through a chat function, to promote long-term compliance with the intervention \cite{Bickmore2005EstablishingRelationships}. RehabCoach is not meant to deliver therapy directly but takes advantage of MobileCoach’s digital coaching system, for instance to send messages to encourage patients to adhere to predefined therapy plans or to give feedback, instead of only implementing conventional reminders \cite{Jamieson2017, Micallef2016}. MobileCoach-based interventions have already been applied and validated in different populations and applications \cite{Kowatsch2021ConversationalStudy, Kramer2020WhichTrial}, but never in people after stroke, who might present additional challenges regarding the use of digital tools.

Our long-term vision is to use the RehabCoach app in combination with one or more robotic devices in order to increase patients’ adherence and motivation to train unsupervised (in the clinic or at home). As an initial step towards this, we developed a first version of RehabCoach to be presented to primary (i.e., stroke patients) and secondary (i.e., healthcare professionals) users in a single evaluation session. This first version is not yet connected to a rehabilitation device, but still allows users to be involved early in a first iteration of the RehabCoach app design, to collect data and feedback on usability to be incorporated into the next version, promoting its acceptance in the future \cite{VanDerLinden2012}. Furthermore, testing the app separately from a robotic device avoids introducing patients to many different technologies at the same time, limiting potential confounding factors when evaluating usability. 

In this paper, we describe the interdisciplinary development and preliminary usability assessment of RehabCoach. The primary goal is to assess whether individuals after a stroke can successfully interact with such a digital tool after minimal instructions. The secondary goal is to collect quantitative and qualitative feedback on the usability of RehabCoach and gather suggestions for improvements from both primary and secondary users, which could help drive the design of future digital interventions supporting rehabilitation after stroke. 

\section{METHODS}

\subsection{RehabCoach design}

The RehabCoach app builds on MobileCoach, an open-source platform for smartphone-based ecological momentary assessments and health interventions that focus on health-promoting behavior and psychoeducation \cite{MobileCoach, Filler2015MobileCoach:Context, Kowatsch2017DesignMobileCoach}. MobileCoach supports the implementation of a chat-based user interface with a conversational agent and pre-defined answer options. The use of a smartphone-based conversational agent in individuals after stroke raises additional challenges, as this population is often not used to technology and may suffer from cognitive and sensorimotor impairments that could hinder the use of such an app.

\begin{figure}[h]
	\includegraphics[width=0.5\linewidth, center]{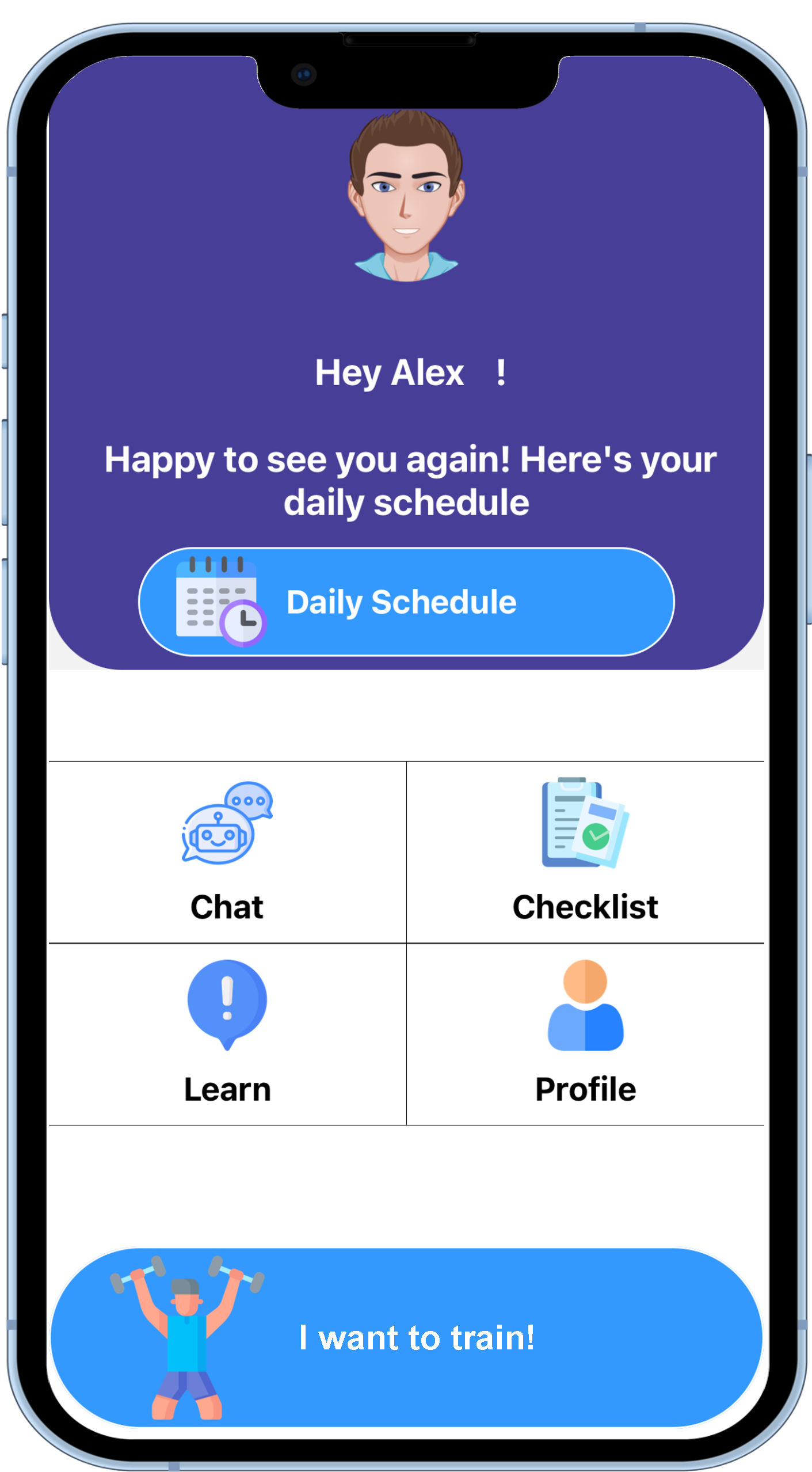}
	\caption{Screenshot of the main menu of the RehabCoach app.}
	\label{Menu_app}
\end{figure}

For the first prototype of RehabCoach (version 0), the existing MobileCoach features previously applied to other populations were combined into a therapy coaching concept, collaboratively discussed by experts in neurorehabilitation, computer science, digital health, and information system. Also, existing literature on mobile apps and means to boost motivation to train for elderly and stroke patients was consulted and combined with prior experience with MobileCoach-based interventions in other populations. This resulted in the definition of a first set of key features relevant to neurorehabilitation and, in particular, to unsupervised therapy (Figure \ref{Menu_app}). 

\textit{\bfseries{Profile:}} When opening the RehabCoach app for the first time, the patient must enter some information (e.g., name, ability to type on the phone, ability to walk) which could then be used to personalize the intervention (i.e., the set of interactions with RehabCoach aimed at supporting unsupervised therapy). The key information to enter was discussed and approved by healthcare professionals. Furthermore, the patient can choose between two avatars both representing a digital coach.     

\textit{\bfseries{Chat:}} The digital coach can send messages to the user via a chat interface. These messages are designed to increase adherence to pre-defined therapy schedules, for example, by reminding the user to start a planned training session. To increase motivation, messages containing feedback about the daily results can be sent \cite{{Oyake2020}}. During the day, the digital coach can also send messages which are not strictly relevant to therapy but relate, for instance, to the emotional or physical state of the user (e.g., “How was your day?”). This aims at strengthening the personal relationship with the user, promoting long-term interaction with RehabCoach \cite{Bickmore2005EstablishingRelationships}, and compensating to a certain extent for the lack of direct social contact during unsupervised therapy.

\textit{\bfseries{Learn:}} Another detrimental factor to patients’ wellbeing is the lack of understanding towards stroke and secondary stroke prevention \cite{Faiz2018}. RehabCoach offers a learning section where videos on stroke, health, and the importance of rehabilitation can be accessed. The goal of this section is to increase patients’ awareness about their disease and health status, which may lead to higher adherence to therapy, the adoption of a healthier lifestyle, and act as a secondary stroke prevention intervention \cite{Croquelois2006, Oyake2020}.

\textit{\bfseries{Checklist:}} The overview of the daily goals (e.g., training sessions) which have already been achieved or still need to be completed is provided in the form of a checklist. Clearly stating the daily goals should increase users’ engagement and, thus, adherence to the therapy schedule \cite{Charles2020}.   

\textit{\bfseries{“I want to train”:}} The virtual coach sends messages related to a training session based on a pre-defined schedule. The “I want to train” feature allows the user to start a training session spontaneously, in addition to the training program. Since unsupervised therapy could lead to patients not training consistently, this feature allows the user of the app to perform a training session even if they skipped a scheduled one, or to train more.

 \begin{figure*}[h]
	\includegraphics[width=\textwidth]{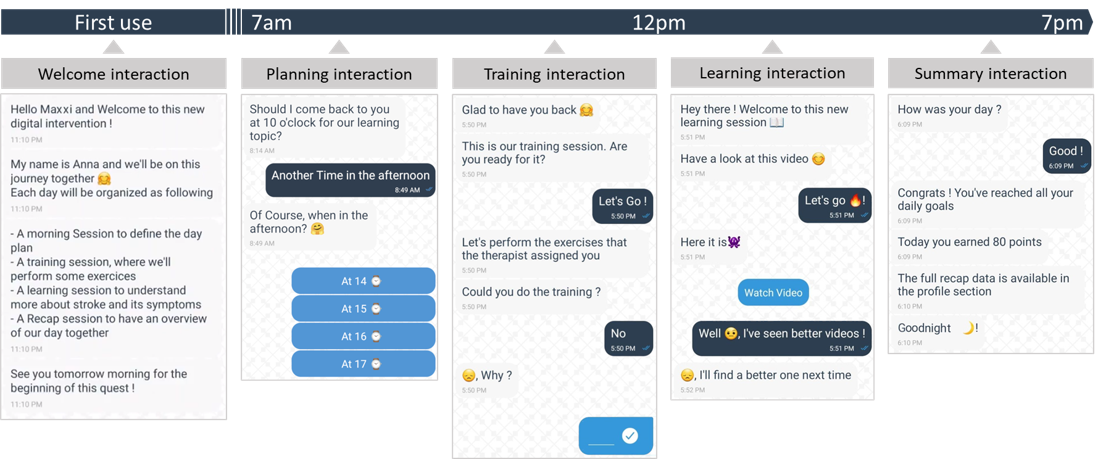}
	\caption{Excerpts from representative chat-based daily interactions between a user and RehabCoach. Grey boxes correspond to the messages sent by the virtual coach, while dark blue boxes are messages sent by the user. Depending on the question, the user can answer either by choosing between predefined answers and clicking on the corresponding button (e.g., light blue buttons with different time options in the planning interaction) or by typing an answer in a predefined field (e.g., light blue box in the training interaction). While the planning interaction and the summary interaction are always at a fixed time (8 am and 7 pm, respectively), the time for the training interaction and the learning interaction can vary depending on the patient’s answers.}
	\label{Daily_interactions}
\end{figure*}

It is envisioned that the features described above are exploited differently throughout a typical day of using the app to support unsupervised therapy (Figure \ref{Daily_interactions}). The first time the user opens the app, RehabCoach explains how the different interactions and sessions work (welcome interaction). Then, every morning, a planning interaction is performed to plan the daily training and learning sessions at a time suitable for the user, depending on their daily schedule. Therefore, during this interaction, the number of training sessions and the time for the training and learning sessions can be set. If the user does not answer the messages sent by the digital coach during this interaction, the sessions are set to a default time. At the set time, the training and learning interactions start with a message from the digital coach reminding the user to perform the therapy exercises or to watch an informative video. Users can either confirm that they will start the session or postpone it. The digital coach then asks for feedback on the session, providing the opportunity to collect subjective information on the quality of a therapy session. Finally, in the evening, RehabCoach summarizes the user’s daily performance (summary interaction). When the app is not open, messages appear on the smartphone as push notifications. If users do not respond within a certain time frame (i.e., 10 minutes), the interaction is tagged as incomplete, and the app proceeds to the next interaction scheduled. All the answers inputted by the user are transmitted over an encrypted channel and saved automatically on a secure web server.

\subsection{Preliminary usability evaluation}

The goal of this preliminary usability evaluation was to identify potential technical or accessibility issues and to gather feedback on the app to decide on its further development before combining it with a rehabilitation device and conducting a larger feasibility study at home. The study protocol was approved by the ethics commission of ETH Zurich (2022-N-29). For each participant, the usability evaluation consisted of a single study session of about 45 minutes, simulating an entire day of using the app. The evaluation included individuals after a stroke and healthcare professionals working with stroke patients. Healthcare professionals were included with the objective to gain additional feedback on the app, for example, on further features that could be implemented in upcoming versions. This group would further serve as a baseline regarding feasibility and performance in the given tasks with the app. To be included in the study, participants had to be older than 18 years and sign the informed consent. An additional inclusion criterion for primary users was a diagnosis of stroke, while healthcare professionals had to regularly interact with individuals post-stroke. Persons with major cognitive and/or communication deficits, or major comprehension and/or memory deficits reported in their clinical records were excluded. The usability evaluation session consisted of three phases:

\textit{\bfseries{Familiarization:}} The researcher described the aim of the project, answered eventual questions, and checked that participants understood and signed the informed consent form. Then, the researcher presented the RehabCoach app on a test smartphone and explained how to interact with it. This was then followed by three minutes of familiarization, where the participant could freely interact with the app and ask questions.

\textit{\bfseries{Testing:}} Participants were asked to perform specific tasks (Figure 3a) considered essential for correct daily interactions with the app. The time taken to complete each task and task success were recorded. Task success could be rated as: successfully completed, completed with an input from the researcher, completed with an error, or not completed.

\textit{\bfseries{Evaluation:}} Participants were asked to fill in the mHealth App Usability Questionnaire (MAUQ, \cite{Zhou2019}) and to rate four custom statements. Both the MAUQ and the custom statements were rated on a seven-point Likert scale ranging from 1 (strongly agree, i.e., positive opinion about the app and high usability) to 7 (strongly disagree), with the possibility of leaving out an item in case it was not applicable (I don’t know). Items 1-5 of the MAUQ are specific to ease of use (MAUQ\_E), items 6-12 to interface and satisfaction (MAUQ\_I), and items 13-18 to usefulness (MAUQ\_U). This was followed by a semi-structured interview that touched topics related to first impressions about RehabCoach, likes and dislikes, additional features that participants would like to have, design, and informative videos. 

\subsection{Data analysis}
Boxplots were used to represent the time taken to complete the tasks for the group of primary and secondary users separately. Given the small sample size of this preliminary study, no statistical testing was reported. Task success was represented with a heat map to visualize tasks potentially causing more difficulties. MAUQ scores were calculated as the mean of the scores given to the single items, excluding the items where participants answered \textit{I don’t know}. 

\section{RESULTS}
Four subacute stroke patients between 45 and 78 years old (mean age: 63.75 years) and five healthcare professionals between 28 and 51 years old (mean age: 40.2 years) took part in the study. Among the healthcare professionals, one was a speech therapist, one a neuropsychologist, two were physiotherapists and one had a double specialty in physiotherapy and neuropsychology.

\subsection{Task success and completion time}
\begin{figure*}[h]
	\includegraphics[width=\textwidth]{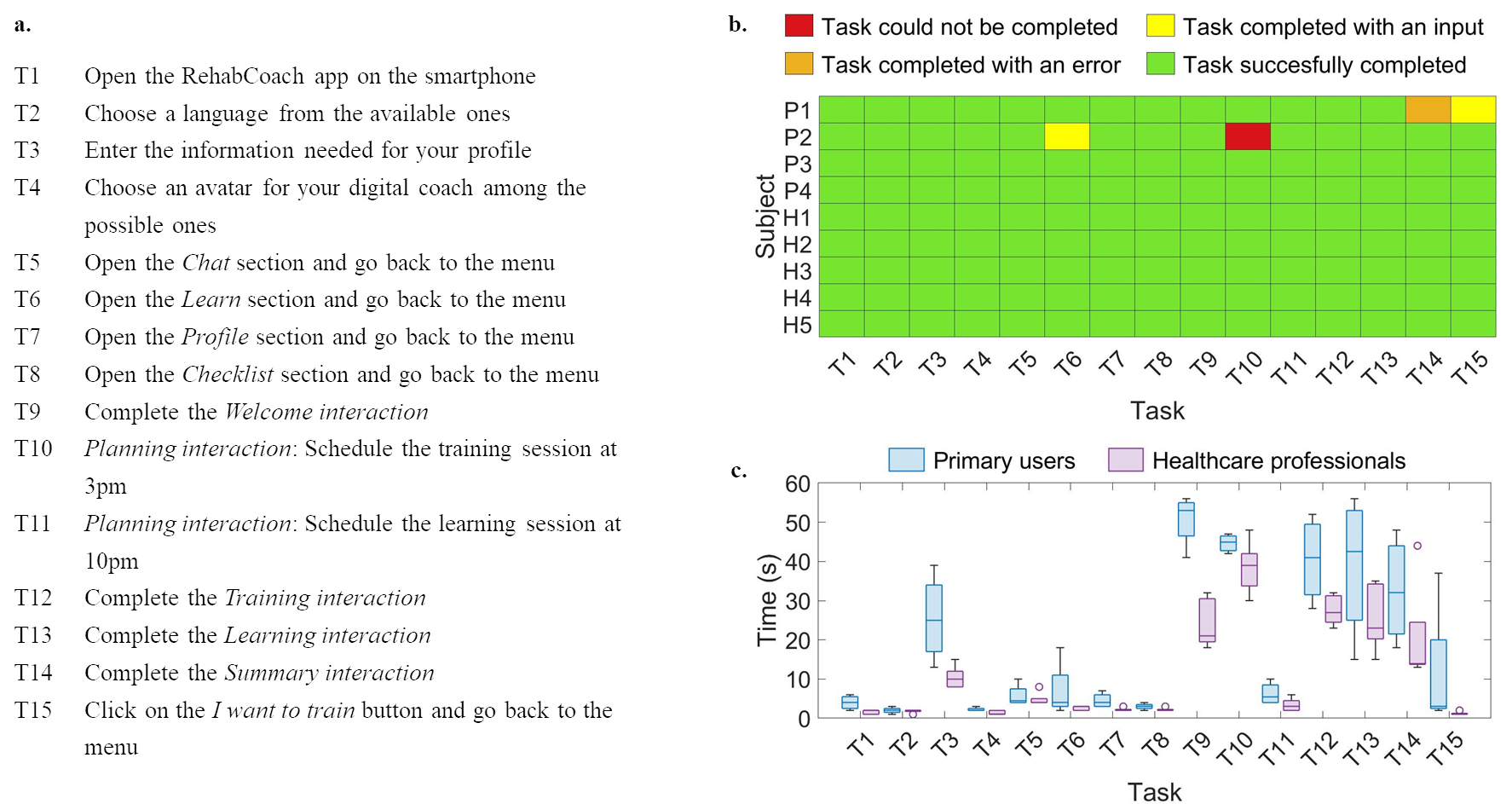}
	\caption{Tasks performed by the participants with the RehabCoach app during the usability evaluation (a), heat map representing task success for both primary users (P1-P4) and healthcare professionals (H1-H5) (b), and boxplots of the time taken to complete the different tasks for the group of primary users and for the one of healthcare professionals (c).}
	\label{Results_all}
\end{figure*}

All healthcare professionals and two of the primary users could complete all tasks (Figure~\ref{Results_all}b). P1 completed the summary interaction (T14) with an error, as an empty message was sent instead of typing an answer to one of the questions. During the planning interaction (T10), P2 accepted the proposed time for the training session (i.e., 2 pm) instead of changing it to 3 pm and therefore did not successfully complete the task. The input given by the researcher (n=2) consisted in encouraging participants to carefully look at one specific section of the app interface to find the icon of interest after the participants expressed difficulties in finding it. For ten tasks, the time taken by the primary users was similar to the healthcare professionals (difference of medians: \textless6 seconds) (Figure~\ref{Results_all}c). However, for Tasks 3, 9, 12, 13, and 14, the time taken by the stroke patients was found to be longer (difference of medians: 14-32 seconds).

\subsection{MAUQ and custom questionnaire}
The mean overall MAUQ score for the primary users was 1.3 (range: 1-1.8), while for the healthcare professionals it was 1.4 (range: 1-1.8). Mean MAUQ\_E scores were 1.1 and 1.6 for primary users and healthcare professionals, respectively, MAUQ\_I scores were 1.3 and 1.4, and MAUQ\_U scores were 1.9 and 1.5. The mean scores assigned to the custom statements are listed in Table~\ref{custom_q}.

\begin{table}[h]
	\caption{Custom statements with mean (range) scores assigned by primary users (P, n = 4) and healthcare professionals (H, n = 5). Possible scores ranged from 1 (strongly agree) to 7 (strongly disagree).}
	\label{custom_q}
	\begin{center}
		\begin{tabular}{l l l}
			\hline	
             \textbf{Statement} & \textbf{P} & \textbf{H} \\
            \hline	
            1: The font size and the colors of the app & 2.3 (1-3) & 2.4 (1-4)\\
                 are adequate &  & \\
			2: The size of the icons is adequate & 1 (1-1) & 2.2 (1-4)\\
			3: The questions asked by the virtual coach & 1 (1-1) & 1.8 (1-4)\\
                 are easy to understand &  &\\
			4: It is easy to answer the questions of the & 1 (1-1) & 1.4 (1-2)\\
                 virtual coach in the chat with the buttons & & \\  
			\hline
		\end{tabular}
	\end{center}
\end{table}

\subsection{Semi-structured interview}
Eight participants specifically stated that RehabCoach is easy to use. Regarding the chat function, a participant mentioned that it humanizes the interaction. One patient mentioned that more choices for the avatar of the digital coach should be provided for increased personalization. Similarly, it was requested to enhance the chat feature so that users can initiate interactions themselves, for instance, by asking common questions such as “How are you?”. One patient mentioned that the empathy of RehabCoach should be improved, for example, by adding interactions that aim at strengthening the attachment bond even more, even though they might not be directly linked to therapy. 

Regarding the chat interface, three participants mentioned that they liked the buttons with predefined answers. Still, two healthcare professionals and one primary user suggested adding more options, especially when the possible answers have opposite meanings and do not cover a broad spectrum of possible responses. For two secondary users more settings should be added to the personalized profile, especially for the chat background, the font size, choosing between capital or lowercase letters, and asking for aphasia. Furthermore, most participants generally liked the idea of having the informative videos. Two healthcare professionals pointed out that the information given should be personalized, for example according to the type of impairment. 

\section{DISCUSSION}
Here we present the first prototype of RehabCoach, a coaching app designed to increase motivation and adherence during unsupervised therapy after stroke, and its preliminary usability testing. RehabCoach is chatbot-based and thought of as a scalable social actor to support and later interface with different devices for unsupervised rehabilitation, depending on the needs of the single patient. Despite the many existing apps for people after stroke \cite{Burns2021}, the potential of chatbot-based apps connected to rehabilitation devices still needs to be investigated. Our results show that primary users could correctly interact with the basic features of RehabCoach after only a short familiarization phase, as demonstrated by high task success and perceived usability. This is aligned with the findings of Burns et al. \cite{Burns2021}, who reviewed 49 articles related to mobile health app interventions for people after stroke, with most of the articles evaluating feasibility concluding that the tested digital interventions were feasible. 

RehabCoach was perceived as user-friendly and the chat messages were easy to understand, which is a critical requirement when designing a digital health intervention for the elderly with potential cognitive impairments, as our target population. The tested tasks were representative of the expected interactions during a day of use of RehabCoach and the different features implemented in this prototype. For most of the tasks, primary users did not take longer than healthcare professionals to complete them. Two out of the five tasks that resulted in noticeably longer execution time by the primary users required typing (T3) or reading long text messages (T9), which is not surprising as literature shows that elderly read and text slower than young adults \cite{Kliegl2004, Krasovsky2018}. This suggests that short messages and predefined answers may be preferred and more suitable for our population. However, in these tasks, the priority was not to be fast but to correctly interact with the app or interpret the information. Therefore, the average additional time needed by primary users to complete all the tasks (+111 seconds in total) is acceptable, also considering that in the intended use case, this additional time would be split over the entire day.

Overall, the MAUQ results were very positive despite the currently limited functionalities of the prototype, which supports the need for developing a more advanced version of RehabCoach in the future. The results of the semi-structured interviews will be used to define key features to implement in the next prototype. For example, while the predefined answers were appreciated as they provided an easier and faster way to answer than typing, three participants stated that the offered predefined answers should be further developed to cover a broader range of possible opinions. Indeed, for most questions we opted to provide only two possible answers, so as to reduce the number of options to read and increase usability. Issues with predefined answer options were also identified in a previous work involving stroke patients interacting with a chatbot \cite{Epalte2020}. 

As the objective was to evaluate as early as possible the feasibility of using the app in persons with stroke and its usability, the interaction with the app was limited to a single short session (representing an accelerated day), and users did not actually engage in the scheduled therapy sessions. This may influence the perception of the app by the participants. While the number of participants in this usability evaluation is limited, it still allowed us to collect essential preliminary data to show feasibility and to improve the app in the next design iteration, which would be tested in a more extensive study including more participants undergoing unsupervised rehabilitation sessions and interacting with the app over multiple days. This will allow defining whether the positive evaluation of this first prototype can translate into high motivation to train and adherence to unsupervised therapy. In conclusion, the positive results gained during this work underline the feasibility of using chatbot-based interventions in individuals after stroke and open the door to the use of RehabCoach as a bridging digital tool to support different devices for unsupervised therapy after stroke.






\section*{ACKNOWLEDGMENT}

The authors would like to thank the participants of the usability evaluation study and the team of the Clinica Hildebrand in Brissago for their support during this work.

\section*{CONFLICT OF INTEREST}

FS and TK are affiliated with the Centre for Digital Health Interventions (CDHI), a joint initiative of the Institute for Implementation Science in Health Care, University of Zurich, the Department of Management, Technology, and Economics at ETH Zurich, and the Institute of Technology Management and School of Medicine at the University of St.Gallen. CDHI is funded in part by CSS, a Swiss health insurer, MTIP, a Swiss investor, and Mavie Next, an Austrian healthcare provider. TK is also a co-founder of Pathmate Technologies, a university spin-off company that creates and delivers digital clinical pathways. However, neither CSS, MTIP, Mavie Next nor Pathmate Technologies were involved in this research.


\end{document}